\documentclass[twocolumn,showpacs,preprintnumbers,amsmath,amssymb,floatfix]{revtex4}
\usepackage{graphicx}
\usepackage{amsmath}
\usepackage{bm}
\usepackage{color}

\newcommand{\bald}[1]{{\bf #1}}

\usepackage{dcolumn}% Align table columns on decimal point

\begin{document}

\title{The sound produced by a fast parton in the quark-gluon plasma is a ``crescendo''}

\author{R. B. Neufeld and B. M\"uller}
   \affiliation{Department of Physics, Duke University, Durham, NC 27708, USA}

\date{\today}

\begin{abstract}

We calculate the total energy deposited into the medium per unit length by fast partons traversing a quark-gluon plasma.  The medium excitation due to collisions is taken to be given by the well known expression for the collisional drag force.  The radiative energy loss of the parton contributes to the energy deposition because each radiated gluon acts as an additional source of collisional energy loss in the medium.  We derive a differential equation which governs how the spectrum of radiated gluons is modified when this energy loss is taken into account.  This modified spectrum is then used to calculate the additional energy deposition due to the interactions of radiated gluons with the medium. Numerical results are presented for the medium response for the case of two energetic back-to-back partons created in a hard interaction.
\end{abstract}


\maketitle

Experimental results from the heavy-ion program at the Relativistic Heavy-Ion Collider (RHIC) indicate the creation of a new state of nuclear matter known as the quark-gluon plasma \cite{Arsene:2004fa}.  The results obtained so far indicate some fascinating properties of the  quark-gluon plasma, not the least of which is the nearly perfect fluid behavior apparently exhibited by the medium \cite{Romatschke:2007mq}.  Another significant observation \cite{Adcox:2001jp} is that highly energetic partons which propagate through the medium rapidly lose energy and momentum as they interact with the surrounding matter in a process known as jet quenching \cite{Wang:1991xy}.  An interesting question related to the above properties is: how does the  quark-gluon plasma respond to fast partons as they propagate through it?  This question has gained significance in light of measurements of hadron correlation functions that are consistent with a conical emission pattern \cite{Adams:2005ph,Adler:2005ee,Ulery:2007zb,Adare:2008cq}.  These measurements suggest that fast partons may generate collective flow, such as a Mach-cone shaped shock wave, in the medium.

While jet quenching has been studied quite extensively \cite{jetquenching,Baier:2000mf,Jacobs:2004qv}, theoretical studies of the quark-gluon plasma response to fast partons are more recent \cite{Machcone,Renk:2005si,Betz:2008js,Neufeld:2008fi,Neufeld:2008hs,Neufeld:2008dx}.  The central challenge of this investigation is to calculate the energy deposited into the medium per unit length by an energetic parton.  As the fast parton propagates through the medium, it loses energy through collisions and medium induced radiation.  
%The collisional energy loss goes directly into medium excitation, however, the radiated gluons still need to deposit their energy.  
The energy deposited into the medium per unit length is then the sum of the collisional energy loss of the primary parton and of the radiated gluons.

In this work we calculate the total energy deposited into the medium per unit length by a fast parton traversing a quark-gluon plasma.  The medium excitation due to collisions is taken to be given by the well-known expression for the collisional drag force \cite{Thoma:1991ea}.  To calculate the medium excitation due to radiation, we begin by deriving a differential equation which describes how the spectrum of radiated gluons is modified as the radiated gluons lose energy through collisions.  This modified spectrum is then used to calculate the differential energy loss due to the interactions of radiated gluons with the medium, from which we find that the energy which goes into medium excitation is substantially less than the total radiative energy loss.  The final result for the energy deposited into the medium per unit length, which is a sum of the primary and the secondary contributions, is then treated as the coefficient of a local hydrodynamic source term.  Numerical results are presented for the medium response for the case of two fast, back-to-back partons created in an initial hard interaction.

We start by considering the collisional energy lost per unit length by a fast parton in the  quark-gluon plasma.  As mentioned above, for the collisional energy loss we will use the familiar expression \cite{Thoma:1991ea}
\begin{equation}\label{dE_dx_c}
\left(\frac{dE}{d \xi}\right)_C = \frac{\alpha_s C_2 \, m_{D}^2}{2}  \ln \frac{2 \sqrt{E_p T}}{m_D}
\end{equation}
where $\alpha_s = g^2/4\pi$ is the strong coupling, $m_{D}$ is the Debye mass of the medium, which we take to be given by $ m_{D} =  gT$, $E_p$ is the energy of the fast parton, $T$ is the temperature of the medium, and $C_2$ is the eigenvalue of the quadratic Casimir operator of the color charge of the source parton, which is 4/3 if the fast parton is a quark, and 3 for a gluon (we consider $N_c = 3$).  In (\ref{dE_dx_c}), the subscript $C$ denotes collisional energy loss and $\xi$ is the path-length traveled by the source parton.

We are next interested in calculating the energy deposited, or gained by the medium, due to gluons radiated by the fast parton.  We begin by defining the quantity $f(\omega,\xi) \equiv d I_M/d\omega$, which gives the spectrum of radiated gluons with energy $\omega$ in the medium. $f(\omega,\xi)$ is in general different than $d I /d \omega$, which is the spectrum of gluons emitted by the fast parton, because gluons, once emitted, lose energy in the medium due to collisions until they become part of the thermal bath.  As a gluon with energy $\omega$ travels from $\xi$ to $\xi + \Delta \xi$, it loses collisional energy $\epsilon(\omega)\, \Delta \xi$, where $\epsilon(\omega)$ is obtained from (\ref{dE_dx_c}) to be
\begin{equation}
\epsilon(\omega) = \frac{3}{2} \alpha_s \, m_{D}^2 \ln \frac{2 \sqrt{\omega T}}{m_D}.
\end{equation}
Thus, in order to find a gluon with energy $\omega$ at position $\xi + \Delta \xi$, there must be a gluon with energy $\omega + \epsilon(\omega)\, \Delta \xi$ at position $\xi$.  Additionally, we require the total number of gluons, that is, $f \, d \omega$, to be invariant.  This means that as $\omega \rightarrow \omega + \epsilon(\omega)\, \Delta \xi$ one has $d \omega \rightarrow d\omega(1 + \frac{\partial \epsilon(\omega)}{\partial \omega} \Delta \xi$).  In equation form, this is
\begin{equation}\label{crude}
f(\omega,\xi + \Delta \xi) = f(\omega + \epsilon(\omega)\, \Delta \xi,\xi)(1 + \frac{\partial \epsilon(\omega)}{\partial \omega} \Delta \xi).
\end{equation}
Furthermore, as the fast parton moves from $\xi$ to $\xi + \Delta \xi$, it will emit additional gluons, $\Delta \xi \times d I/d\omega d\xi$, which add to $f(\omega,\xi+\Delta\xi)$. The evolution equation for $f(\omega,\xi)$ thus takes the form
\begin{equation}\label{diff_e_q}
\frac{\partial}{\partial \xi}f(\omega,\xi) - \frac{\partial}{\partial \omega} \left[\epsilon(\omega)\, f(\omega,\xi)\right] = \frac{d I}{d \omega d \xi}(\omega,\xi)
\end{equation}
where we have taken the limit of $\Delta \xi \rightarrow 0$.

Equation (\ref{diff_e_q}) provides a partial differential equation through which one can determine the spectrum of radiated gluons in the medium, $f(\omega,\xi)$.  In order to solve for $f$, it is necessary to specify $dI/d\omega d\xi$.  We choose the spectrum calculated by Salgado and Wiedemann in the multiple soft scattering approximation \cite{Salgado:2003gb}
\begin{equation}\label{specchoice}
\frac{d I}{d \omega d \xi} = - \frac{\sqrt{\hat{q}} \, \alpha_s C_2}{\pi} \text{Re} \frac{(1 + i)\tan \left[(1+i)\sqrt{\frac{\hat{q}}{\omega}} \frac{\xi}{2}\right]}{\omega^{3/2}}
\end{equation}
where we use \cite{Baier:2008zz}
\begin{equation}\label{qhat}
\hat{q} = 2 \alpha_s C_2 m_D^2 T \ln \frac{2 \sqrt{E_p T}}{m_D},
\end{equation}
where the logarithm is consistent with (\ref{dE_dx_c}).

\begin{figure*}
\centerline{
\includegraphics[width = 0.46\linewidth]{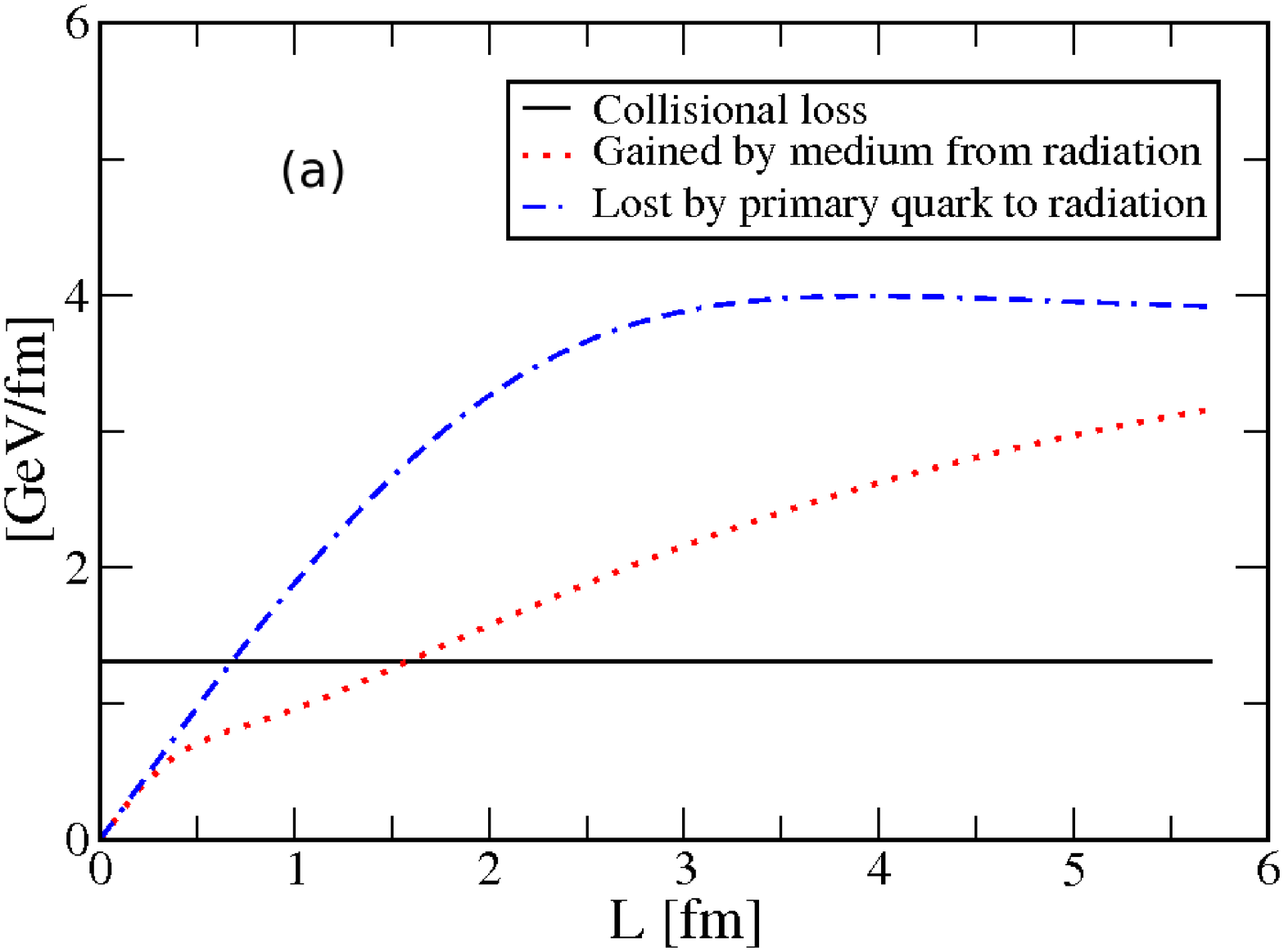}
\includegraphics[width = 0.46\linewidth]{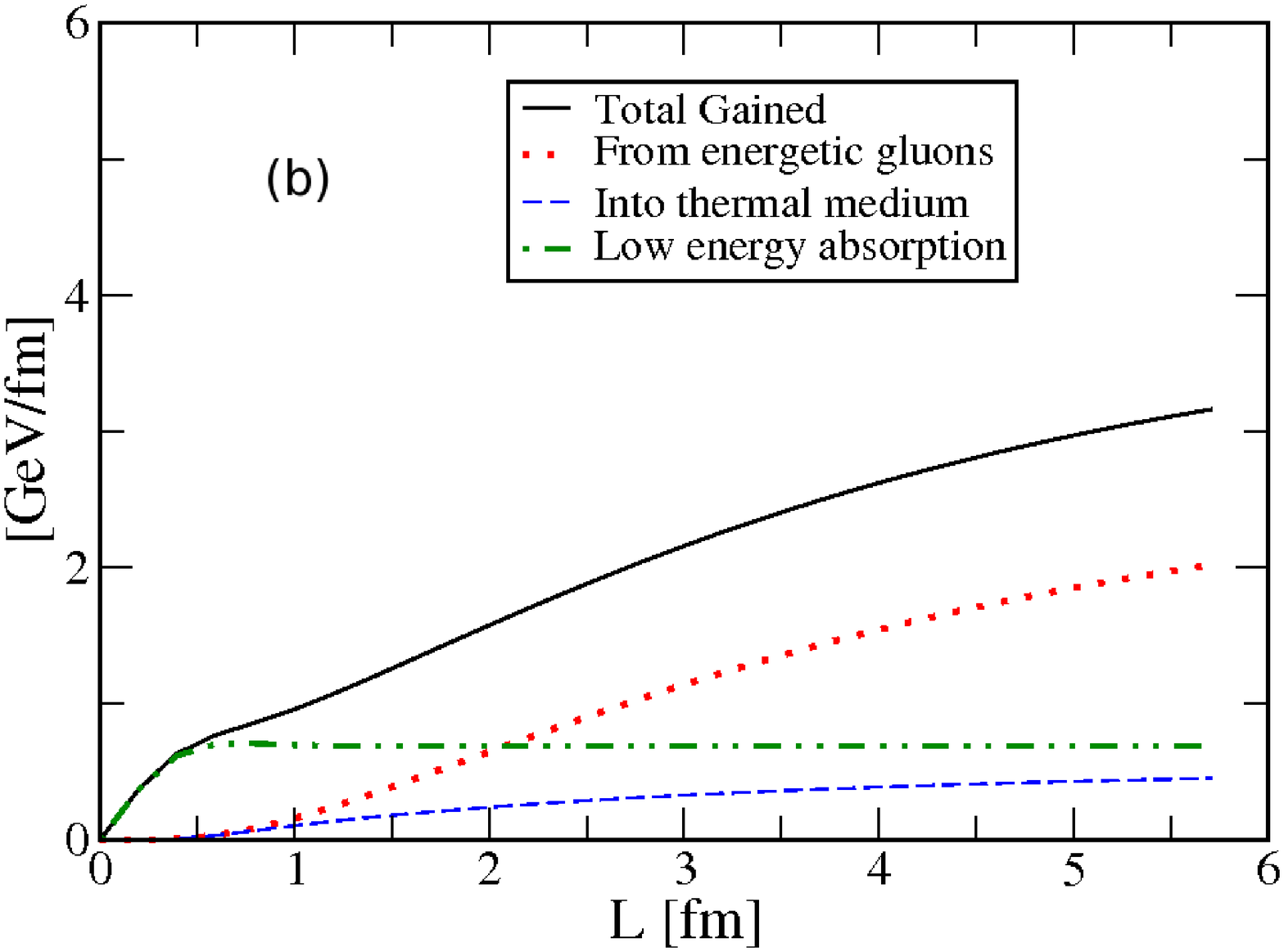}
}
\caption{(Color online) Plot (a) shows the differential energy lost by the fast parton due to radiation and that gained by the medium as a function of path-length, $L$, as well as the collisional energy loss (the specific parameters are discussed in the text). The differential energy deposition into the medium is the sum of the solid (black) and dotted (red) lines. Plot (b) shows the breakdown of various the contributions to the radiative energy gained by the medium.  The dotted (red) line shows the collisional energy loss of the radiated gluons. The dash-dotted (green) line shows the the energy radiated in gluons with energy below $\bar\omega$, which are assumed to immediately become part of the medium. The (blue) curve marked 'Into thermal medium' denotes the last term in (\ref{dE_dx_r}). 
}
\label{total_energy}
\end{figure*}

$f(\omega,\xi)$ will in general consist of two components: energetic gluons which lose energy through collisions and low energy gluons which become a part of the medium.  The total energy being dumped into the medium is then given by the sum of the collisional energy loss of high energy gluons and the energy absorbed by the medium in the form of low energy gluons.  We make the distinction that gluons with $\omega > \bar{\omega} \equiv 2 T$ are sources of collisional energy loss, while those with less energy are considered as immediately part of the medium.  For $\omega > \bar{\omega}$ we solve for $f(\omega,\xi)$ numerically from equation (\ref{diff_e_q}) for a primary quark using the parameters: $\alpha_s = 1/\pi$, $T = 350$ MeV, and $E_p = 50$ GeV.  The total energy deposited into the medium by the secondary gluons per unit length is then given by
\begin{equation}\label{dE_dx_r}
\begin{split}
\left(\frac{dE}{d \xi}\right)_R = \int_{\bar{\omega}}^{\omega_{\rm max}} d \omega \, &\epsilon(\omega) \, f(\omega,\xi) + \int_{0}^{\bar{\omega}} d \omega \, \omega \frac{d I}{d \omega d \xi} \\
& + \bar{\omega} \, f(\bar{\omega},\xi)\, \epsilon(\bar{\omega}),
\end{split}
\end{equation}
where we set $\omega_{\rm max} = E_p/2$.  The last term in equation (\ref{dE_dx_r}) accounts for energetic gluons which have lost enough energy to become a part of the medium and thus deposit their entire remaining energy.  When solving (\ref{diff_e_q}), we multiply the spectrum (\ref{specchoice}) by a factor of $1 - (\omega/\omega_{\rm max})^8$ to ensure it goes to zero at $\omega = \omega_{\rm max}$.

The result of (\ref{dE_dx_r}) is shown as the dotted red line in Fig.~\ref{total_energy}(a) for the same parameters listed above, along with the differential collisional energy loss of the primary parton (solid black line) and its differential energy loss to radiation (dashed blue curve).  One sees that the energy deposited by the radiation into the medium per unit length has an approximately linear growth in pathlength, which results from the steady increase in the number of gluons that deposit energy in collisions.  This linear growth is thus of a different origin than what is observed in the first few fm of the differential energy loss to radiation, which is caused by the energy dependent coherence length for radiated gluons.  If one continues the curves shown in Fig.~\ref{total_energy}(a) out to a large enough pathlength, the dotted red line reaches a steady state solution and merges with the dashed blue curve.  The individual contributions on the right-hand side of (\ref{dE_dx_r}) are shown separately in Fig.~\ref{total_energy}(b), together with their sum.

The total energy deposited into the medium per unit length, or time, is given by the sum of (\ref{dE_dx_c}) and (\ref{dE_dx_r}), which we write as
\begin{equation}\label{tot_diff_loss}
\frac{d E}{d t} = \left(\frac{dE}{d t}\right)_C + \left(\frac{dE}{d t}\right)_R
\end{equation}
We treat the fast parton as a point source of energy and momentum deposition in the medium, with velocity $\bald{u}$ and energy $E_p$. The hydrodynamic source term, denoted as $J^\nu$, gives the energy and momentum density deposited in the medium per unit time.  For a relativistic point source, $J^\nu$ takes the following form
\begin{equation}\label{simp_source}
\begin{split}
J^\nu(x) &= \frac{dE}{dt}\delta(\bald{x} - \bald{u} t)\left(1,\bald{u}\right)
\end{split}
\end{equation}
where $dE/dt$ is given by (\ref{tot_diff_loss}).  To make the calculation more tractable, we fit the result of (\ref{dE_dx_r}) to a linear function of time, from which we find
\begin{equation}
\left(\frac{dE}{d t}\right)_R \approx 0.474 \frac{\text{GeV}}{\text{fm}} + 0.51 \, t \frac{\text{GeV}}{\text{fm}^2} .
\end{equation}
The fit slightly overestimates the energy deposition rate for $t<0.5$ fm/$c$.

The source term couples to the hydrodynamic equations for the medium, $\partial_\mu T^{\mu \nu} = J^{\nu}$, where $T^{\mu \nu}$ is the energy-momentum tensor.  For simplicity, we here consider the source to be coupled to an otherwise static medium at temperature $T$.  Furthermore, we make the assumption that the energy and momentum density deposited by the fast parton is small compared to the equilibrium energy density of the medium, in which case the hydrodynamical equations can be linearized.

We are here interested in calculating the energy density perturbation excited in the medium, denoted as $\delta \varepsilon \equiv \delta T^{00}$.  The equations of motion for a medium coupled to a source in linearized hydrodynamics are discussed in several places (for instance, \cite{Neufeld:2008dx}).  We here write down the result for $\delta \varepsilon$ in Fourier space, which, to first order in the ratio of shear viscosity to entropy density, $\eta/s$, is given by
\begin{equation}\label{eps}
\delta\varepsilon (x) = \int \frac{d^4 k}{(2 \pi)^4}  \,e^{- i k \cdot x} \frac{i k J_L(k)  + J^0(k)(i \omega -  \Gamma_s k^2)}{\omega^2 -  c_s^2 k^2 + i \Gamma_s \omega k^2}
\end{equation}
where $c_s$ denotes the speed of sound, $\Gamma_s \equiv 4 \eta/(3 s T)$ is the sound attenuation length, and the source vector has been divided into transverse and longitudinal parts: ${\mathbf J} = \hat{\mathbf k} J_L + {\mathbf J}_T$, with $\hat{\mathbf k}$ denoting the unit vector in the direction of ${\mathbf k}$.  We here consider two fast partons created in an initial hard interaction at time $t = 0$, which then propagate back to back for some time, $t_M$, before being absorbed by the medium.  In this case (\ref{simp_source}) is modified to
\begin{equation}
\begin{split}\label{mod_source}
J^\nu(x) &= \frac{dE}{dt}(\Theta(t) - \Theta(t - t_M)) \times \\
&\left[\delta(\bald{x} - \bald{u} t)\left(1,\bald{u}\right) + \delta(\bald{x} + \bald{u} t)\left(1,- \bald{u}\right)\right]
\end{split}
\end{equation}
where $\Theta(t)$ is 1 if $t>0$ and zero otherwise.  The Fourier transform of (\ref{mod_source}) can be obtained in a straightforward way to obtain an expression for (\ref{eps}).

\begin{figure*}
\centerline{
\includegraphics[width = 0.85\linewidth]{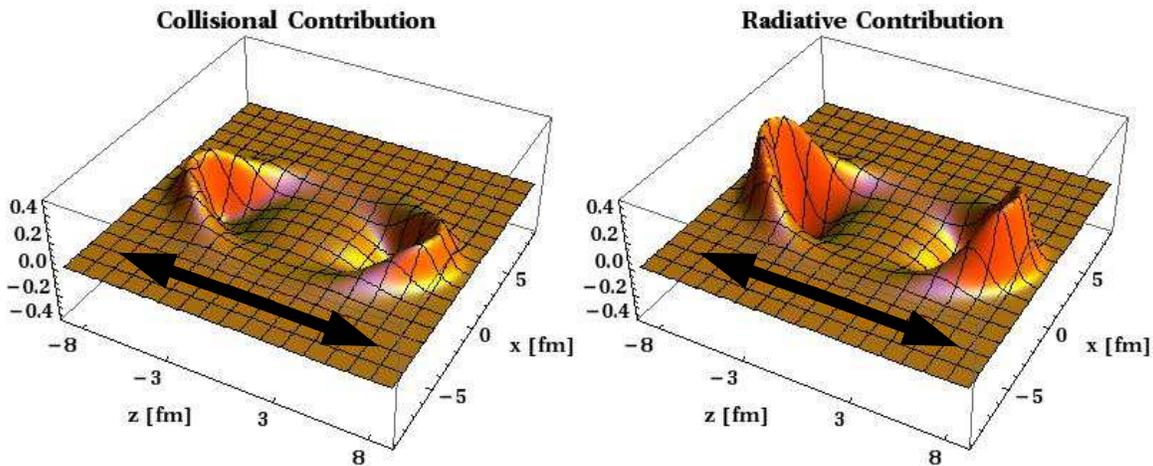}
}
\caption{(Color online) Result for the energy density wave ($\text{GeV}/\text{fm}^3$) excited by back to back quarks propagating along the $\hat{z}$ axis (as indicated by the black arrows).  The plots, which show the collisional and radiative contributions separately, are shown in the $x$-$z$ plane, however, the results are cylindrically symmetric about the $\hat{z}$ axis.  The Mach cone formation is visible in both contributions.  The radiative induced excitation leads to a $t$ growth in the source strength, which can be seen from the plot.
}
\label{mach_wave}
\end{figure*}

The final result for $\delta \varepsilon(x)$ is a sum of contributions from the collisional and radiative energy dumped into the medium.  We calculate the medium response at a time $t = 8$ fm for back to back quarks which are created at $t = 0$ and $\bald{x} = 0$, and propagate along the $\hat{z}$ axis until $t = 6$ fm.  Additionally, each quark is assumed to have an energy $E_p = 50$ GeV.  The result from the collisional and radiative contributions are presented separately in Fig.~\ref{mach_wave}, for a medium with the same parameters used above as well as $\eta/s = 0.2$ and $c_s/c = 0.57$.  The plot shows the energy density wave excited by the source quarks in the $x$-$z$ plane, however, the results are cylindrically symmetric about the $\hat{z}$ axis.  The Mach cone formation is visible in both the collisional and radiative contributions.  The radiative contribution to the source term, (\ref{mod_source}), exhibits a linear growth with time, which is reflected in the shape of the resulting energy density wave.

In summary, we have derived an expression for the total energy deposited into the medium per unit length (or time) by a fast parton propagating in the  quark-gluon plasma, including both collisional and radiative energy deposition.  We have shown that the contribution of gluon radiation to the medium excitation grows with path length.  Our result is reminiscent of, but less dramatic than, the increase of the energy deposition by a light quark obtained in the strongly coupled supersymmetric Yang-Mills theory \cite{Chesler:2008uy}.

In our formalism, the magnitude of the wave depends on the specific value (\ref{qhat}) of $\hat{q}$, as well as the form of the collisional energy loss (\ref{dE_dx_c}).  It is possible that in the quark-gluon plasma produced at RHIC either one or both of these has a larger value than that assumed here (see Bass {\em et al.} \cite{Bass:2008rv} for a range of values of $\hat{q}$ compatible with experimental data).  A general feature of our result is that the perturbation created in the QCD medium by a fast parton will be dominated by the last stage of the phase in which color charges are deconfined and highly mobile, {\em i.~e.} just before bulk hadronization. This observation may explain why the measured peak emission angle of secondary hadrons in the backward direction \cite{Adler:2005ee} corresponds to a small sound velocity $c_s/c \approx 0.3$ consistent with values deduced from lattice QCD for $T\approx T_c$.  

One may ask if the rapid expansion of the medium in a heavy-ion collision may result in decreasing rate of collisional energy loss that counteracts the growth in time of the energy deposition found here.  The collisional energy loss (\ref{dE_dx_c}) is proportional to $g(T)^2m_D^2$.  An examination of lattice results for $m_D$ \cite{Karsch:2006sf} shows that this quantity is almost independent of $T$ in the range relevant to the RHIC experiments. This suggests that the expansion of the medium in a heavy-ion collision will not substantially alter our conclusions.

{\it Acknowledgments:} This work was supported in part by the U.~S.~Department of Energy under grant DE-FG02-05ER41367. We acknowledge valuable discussions with A.~Majumder.  We draw attention to a more recent preprint \cite{Qin:2009uh}, in which Qin {\em et al.} treated the same problem in a different formalism and obtained similar results.


\begin{thebibliography}{99}

%\cite{Arsene:2004fa}
\bibitem{Arsene:2004fa}
  I.~Arsene {\it et al.}  [BRAHMS Collaboration],
  %``Quark Gluon Plasma an Color Glass Condensate at RHIC? The perspective from
  %the BRAHMS experiment,''
  Nucl.\ Phys.\  A {\bf 757}, 1 (2005);
  %[arXiv:nucl-ex/0410020].
  %%CITATION = NUPHA,A757,1;%%
  B.~B.~Back {\it et al.},
  %``The PHOBOS perspective on discoveries at RHIC,''
  Nucl.\ Phys.\  A {\bf 757}, 28 (2005);
  %[arXiv:nucl-ex/0410022].
  %%CITATION = NUPHA,A757,28;%%
  J.~Adams {\it et al.}  [STAR Collaboration],
  %``Experimental and theoretical challenges in the search for the quark  gluon
  %plasma: The STAR collaboration's critical assessment of the  evidence from
  %RHIC collisions,''
  Nucl.\ Phys.\  A {\bf 757}, 102 (2005);
  %[arXiv:nucl-ex/0501009].
  %%CITATION = NUPHA,A757,102;%%
  K.~Adcox {\it et al.}  [PHENIX Collaboration],
  %``Formation of dense partonic matter in relativistic nucleus nucleus
  %collisions at RHIC: Experimental evaluation by the PHENIX  collaboration,''
  Nucl.\ Phys.\  A {\bf 757}, 184 (2005).
  %[arXiv:nucl-ex/0410003].
  %%CITATION = NUPHA,A757,184;%%

\bibitem{Romatschke:2007mq}
  P.~Romatschke and U.~Romatschke,
  %``Viscosity Information from Relativistic Nuclear Collisions: How Perfect is
  %the Fluid Observed at RHIC?,''
  Phys.\ Rev.\ Lett.\  {\bf 99}, 172301 (2007).
  %[arXiv:0706.1522 [nucl-th]].
  %%CITATION = PRLTA,99,172301;%%

%\cite{Adcox:2001jp}
\bibitem{Adcox:2001jp}
  K.~Adcox {\it et al.}, 
  %[PHENIX Collaboration],
  %``Suppression of hadrons with large transverse momentum in central  Au + Au
  %collisions at s**(1/2)(N N) = 130-GeV,''
  Phys.\ Rev.\ Lett.\  {\bf 88}, 022301 (2002);
  %[arXiv:nucl-ex/0109003].
  %%CITATION = PRLTA,88,022301;%%
%\cite{Adler:2002xw}
%\bibitem{Adler:2002xw}
  C.~Adler {\it et al.},  
  %[STAR Collaboration],
  %``Centrality dependence of high p(T) hadron suppression in Au + Au collisions
  %at s(NN)**(1/2) = 130-GeV,''
  Phys.\ Rev.\ Lett.\  {\bf 89}, 202301 (2002).
  %%CITATION = PRLTA,89,202301;%%

%\cite{Wang:1991xy}
\bibitem{Wang:1991xy}
  X.~N.~Wang and M.~Gyulassy,
  %``Gluon shadowing and jet quenching in A + A collisions at s**(1/2) =
  %200-GeV,''
  Phys.\ Rev.\ Lett.\  {\bf 68}, 1480 (1992).
  %%CITATION = PRLTA,68,1480;%%

\bibitem{Adams:2005ph}
  J.~Adams {\it et al.},
  %  [STAR Collaboration],
  %``Distributions of charged hadrons associated with high transverse  momentum
  %particles in p p and Au + Au collisions at s(NN)**(1/2) =  200-GeV,''
  Phys.\ Rev.\ Lett.\  {\bf 95}, 152301 (2005).
  %[arXiv:nucl-ex/0501016].
  %%CITATION = PRLTA,95,152301;%%

\bibitem{Adler:2005ee}
  S.~S.~Adler {\it et al.},
  %  [PHENIX Collaboration],
  %``Modifications to di-jet hadron pair correlations in Au + Au collisions  at
  %s(NN)**(1/2) = 200-GeV,''
  Phys.\ Rev.\ Lett.\  {\bf 97}, 052301 (2006).
  %[arXiv:nucl-ex/0507004].
  %%CITATION = PRLTA,97,052301;%%

%\cite{Ulery:2007zb}
\bibitem{Ulery:2007zb}
  J.~G.~Ulery  [for the STAR Collaboration],
  %``Are There Mach Cones in Heavy Ion Collisions? Three-Particle   Correlations
  %from STAR,''
  Int.\ J.\ Mod.\ Phys.\  E {\bf 16}, 2005 (2007).
  %[arXiv:0704.0224 [nucl-ex]].
  %%CITATION = IMPAE,E16,2005;%%

%\cite{Adare:2008cq}
\bibitem{Adare:2008cq}
  A.~Adare {\it et al.},
  %  [PHENIX Collaboration],
  %``Dihadron azimuthal correlations in Au+Au collisions at sqrt(s_NN)=200
  %GeV,''
  Phys.\ Rev.\  C {\bf 78}, 014901 (2008).
  %[arXiv:0801.4545 [nucl-ex]].
  %%CITATION = PHRVA,C78,014901;%%

\bibitem{jetquenching}
%\cite{Baier:1996sk}
%\bibitem{Baier:1996sk}
  R.~Baier, Y.~L.~Dokshitzer, A.~H.~Mueller, S.~Peigne and D.~Schiff,
  %``Radiative energy loss and p(T)-broadening of high energy partons in
  %nuclei,''
  Nucl.\ Phys.\  B {\bf 484}, 265 (1997);
  %[arXiv:hep-ph/9608322].
  %%CITATION = NUPHA,B484,265;%%
%\cite{Zakharov:1997uu}
%\bibitem{Zakharov:1997uu}
  B.~G.~Zakharov,
  %``Radiative energy loss of high energy quarks in finite-size nuclear  matter
  %and quark-gluon plasma,''
  JETP Lett.\  {\bf 65}, 615 (1997);
  %[arXiv:hep-ph/9704255].
  %%CITATION = JTPLA,65,615;%%
%\cite{Gyulassy:2000fs}
%\bibitem{Gyulassy:2000fs}
  M.~Gyulassy, P.~Levai and I.~Vitev,
  %``Non-Abelian energy loss at finite opacity,''
  Phys.\ Rev.\ Lett.\  {\bf 85}, 5535 (2000);
  %[arXiv:nucl-th/0005032].
  %%CITATION = PRLTA,85,5535;%%
%\cite{Guo:2000nz}
%\bibitem{Guo:2000nz}
  X.~f.~Guo and X.~N.~Wang,
  %``Multiple scattering, parton energy loss and modified fragmentation
  %functions in deeply inelastic e A scattering,''
  Phys.\ Rev.\ Lett.\  {\bf 85}, 3591 (2000);
  %[arXiv:hep-ph/0005044].
  %%CITATION = PRLTA,85,3591;%%
%\cite{Armesto:2004ud}
%\bibitem{Armesto:2004ud}
  N.~Armesto, C.~A.~Salgado and U.~A.~Wiedemann,
  %``Relating high-energy lepton hadron, proton nucleus and nucleus nucleus
  %collisions through geometric scaling,''
  Phys.\ Rev.\ Lett.\  {\bf 94}, 022002 (2005).
  %[arXiv:hep-ph/0407018].
  %%CITATION = PRLTA,94,022002;%%
  
\bibitem{Baier:2000mf}
  R.~Baier, D.~Schiff and B.~G.~Zakharov,
  %``Energy loss in perturbative QCD,''
  Ann.\ Rev.\ Nucl.\ Part.\ Sci.\  {\bf 50}, 37 (2000).
  %[arXiv:hep-ph/0002198].
  %%CITATION = ARNUA,50,37;%%

\bibitem{Jacobs:2004qv}
  P.~Jacobs and X.~N.~Wang,
  %``Matter in extremis: Ultrarelativistic nuclear collisions at RHIC,''
  Prog.\ Part.\ Nucl.\ Phys.\  {\bf 54}, 443 (2005).
  %[arXiv:hep-ph/0405125].
  %%CITATION = PPNPD,54,443;%%

\bibitem{Machcone}
%\bibitem{CasalderreySolana:2004qm}
  J.~Casalderrey-Solana, E.~V.~Shuryak and D.~Teaney,
  %``Conical flow induced by quenched QCD jets,''
  J.\ Phys.\ Conf.\ Ser.\  {\bf 27}, 22 (2005);
  %[Nucl.\ Phys.\  A {\bf 774}, 577 (2006)]
  %[arXiv:hep-ph/0411315].
  %%CITATION = NUPHA,A774,577;%%
%\cite{Stoecker:2004qu}
%\bibitem{Stoecker:2004qu}
  H.~St\"ocker,
  %``Collective Flow signals the Quark Gluon Plasma,''
  Nucl.\ Phys.\  A {\bf 750} (2005) 121;
 %[arXiv:nucl-th/0406018].
  %%CITATION = NUPHA,A750,121;%%
%\bibitem{Chaudhuri:2005vc}
  A.~K.~Chaudhuri and U.~Heinz,
  %``Effect of jet quenching on the hydrodynamical evolution of QGP,''
  Phys.\ Rev.\ Lett.\  {\bf 97}, 062301 (2006);
  %[arXiv:nucl-th/0503028].
  %%CITATION = PRLTA,97,062301;%%
%\cite{Satarov:2005mv}
%\bibitem{Satarov:2005mv}
  L.~M.~Satarov, H.~St\"ocker and I.~N.~Mishustin,
  %``Mach shocks induced by partonic jets in expanding quark-gluon plasma,''
  Phys.\ Lett.\  B {\bf 627} (2005) 64;
  %[arXiv:hep-ph/0505245].
  %%CITATION = PHLTA,B627,64;%%
%\bibitem{Ruppert:2005uz}
  J.~Ruppert and B.~M\"uller,
  %``Waking the colored plasma,''
  Phys.\ Lett.\  B {\bf 618}, 123 (2005).
  %[arXiv:hep-ph/0503158].
  %%CITATION = PHLTA,B618,123;%%

%\cite{Renk:2005si}
\bibitem{Renk:2005si}
  T.~Renk and J.~Ruppert,
  %``Mach cones in an evolving medium,''
  Phys.\ Rev.\  C {\bf 73} (2006) 011901;
 %%CITATION = PHRVA,C73,011901;%%
  Phys.\ Lett.\  B {\bf 646} (2007) 19;
  %%CITATION = PHLTA,B646,19;%%
  %``Three-particle azimuthal correlations and Mach shocks,''
  Phys.\ Rev.\  C {\bf 76} (2007) 014908;
  %[arXiv:hep-ph/0702102].
  %%CITATION = PHRVA,C76,014908;%%
 Int.\ J.\ Mod.\ Phys.\  E {\bf 16} (2008) 3100.
  %%CITATION = IMPAE,E16,3100;%%

%\cite{Betz:2008js}
\bibitem{Betz:2008js}
  B.~Betz {\em et al.}
  %B.~Betz, M.~Gyulassy, D.~H.~Rischke, H.~St\"ocker and G.~Torrieri,
  %``Jet Propagation and Mach Cones in (3+1)d Ideal Hydrodynamics,''
  J.\ Phys.\ G {\bf 35}, 104106 (2008).
  %%CITATION = JPHGB,G35,104106;%%
  %B.~Betz, M.~Gyulassy, J.~Noronha and G.~Torrieri,
  %``Anomalous Conical Di-jet Correlations in pQCD vs AdS/CFT,''
  arXiv:0807.4526 [hep-ph].
  %%CITATION = ARXIV:0807.4526;%%

%\cite{Neufeld:2008fi}
\bibitem{Neufeld:2008fi}
  R.~B.~Neufeld, B.~M\"uller and J.~Ruppert,
  %``Sonic Mach Cones Induced by Fast Partons in a Perturbative Quark-Gluon
  %Plasma,''
  Phys.\ Rev.\  C {\bf 78}, 041901 (2008).
  %[arXiv:0802.2254 [hep-ph]].
  %%CITATION = PHRVA,C78,041901;%%

%\cite{Neufeld:2008hs}
\bibitem{Neufeld:2008hs}
  R.~B.~Neufeld,
  %``Fast Partons as a Source of Energy and Momentum in a Perturbative
  %Quark-Gluon Plasma,''
  Phys.\ Rev.\  D {\bf 78}, 085015 (2008).
  %[arXiv:0805.0385 [hep-ph]].
  %%CITATION = PHRVA,D78,085015;%%

%\cite{Neufeld:2008dx}
\bibitem{Neufeld:2008dx}
  R.~B.~Neufeld,
  %``Propagating Mach Cones in a Viscous Quark-Gluon Plasma,''
  Phys.\ Rev.\  C {\bf 79}, 054909 (2009).
  %%CITATION = ARXIV:0807.2996;%%

\bibitem{Thoma:1991ea}
  M.~H.~Thoma,
  %``Collisional energy loss of high-energy jets in the quark gluon plasma,''
  Phys.\ Lett.\  B {\bf 273}, 128 (1991).
  %%CITATION = PHLTA,B273,128;%%

%\cite{Salgado:2003gb}
\bibitem{Salgado:2003gb}
  C.~A.~Salgado and U.~A.~Wiedemann,
  %``Calculating quenching weights,''
  Phys.\ Rev.\  D {\bf 68}, 014008 (2003).
  %[arXiv:hep-ph/0302184].
  %%CITATION = PHRVA,D68,014008;%%

%\cite{Baier:2008zz}
\bibitem{Baier:2008zz}
  R.~Baier and Y.~Mehtar-Tani,
  %``Jet Quenching And Broadening: The Transport Coefficient Q In An Anisotropic
  %Plasma,''
  Phys.\ Rev.\  C {\bf 78}, 064906 (2008).
  %%CITATION = PHRVA,C78,064906;%%

\bibitem{Bass:2008rv}
  S.~A.~Bass, C.~Gale, A.~Majumder, C.~Nonaka, G.~Y.~Qin, T.~Renk and J.~Ruppert,
  %``Systematic Comparison of Jet Energy-Loss Schemes in a realistic
  %hydrodynamic medium,''
  Phys.\ Rev.\  C {\bf 79}, 024901 (2009).
  %[arXiv:0808.0908 [nucl-th]].
  %%CITATION = ARXIV:0808.0908;%%

%\cite{Chesler:2008uy}
\bibitem{Chesler:2008uy}
  P.~M.~Chesler, K.~Jensen, A.~Karch and L.~G.~Yaffe,
  %``Light quark energy loss in strongly-coupled N = 4 supersymmetric Yang-Mills
  %plasma,''
  arXiv:0810.1985 [hep-th].
  %%CITATION = ARXIV:0810.1985;%%

%\cite{Karsch:2006sf}
\bibitem{Karsch:2006sf}
  F.~Karsch,
  %``Properties of the Quark Gluon Plasma: A lattice perspective,''
  Nucl.\ Phys.\  A {\bf 783}, 13 (2007).
  %%CITATION = NUPHA,A783,13;%%


%\cite{Qin:2009uh}
\bibitem{Qin:2009uh}
  G.~Y.~Qin, A.~Majumder, H.~Song and U.~Heinz,
  %``Energy and momentum deposited into a QCD medium by a jet shower,''
  arXiv:0903.2255 [nucl-th].
  %%CITATION = ARXIV:0903.2255;%%
  
\end{thebibliography}
\end{document}